\documentclass[journal]{article}                                    

\usepackage{arxiv}
\usepackage{hyperref}
\usepackage{graphicx}          
\usepackage{amsmath} 
\usepackage{amssymb}  

\usepackage{amsthm}
\usepackage{color,soul}
\usepackage{flushend}
\usepackage{cite}
\usepackage{mathtools}
\usepackage{xcolor}
\usepackage{array,multirow}
\usepackage{algorithm} 
\usepackage{algorithmic}  
\usepackage[linesnumbered,algo2e,ruled,vlined,norelsize]{algorithm2e}

\usepackage{color,soul}
\usepackage{flushend}
\usepackage{cite}
\usepackage{mathtools}
\usepackage{mfirstuc}
\usepackage{xcolor}
\usepackage{nicematrix}
\usepackage{array,multirow}
\usepackage[ruled,linesnumbered]{algorithm2e}
\usepackage{soul}
\usepackage{multirow}
\usepackage[utf8]{inputenc}
\usepackage{svg}
\usepackage[normalem]{ulem}
\usepackage{adjustbox}

\allowdisplaybreaks
\usepackage{mfirstuc}
\allowdisplaybreaks

\usepackage{authblk}
\title{\LARGE \bf
KAN-Therm: A Lightweight Battery Thermal Model Using Kolmogorov-Arnold Network
} 

\author[1]{Soumyoraj Mallick $^{\star,}$}
\author[1]{Faysal Ahamed $^{\star,}$}
\author[1]{Sanchita Ghosh $^{\star,}$}
\author[1]{Tanushree Roy}
\affil[1]{Department of  Mechanical Engineering, Texas Tech University, Lubbock, TX 79409, US. Emails:~{\tt\small somallic@ttu.edu, fahamed@ttu.edu, sancghos@ttu.edu, tanushree.roy@ttu.edu}.}
\begin{document}

\def\thefootnote{*}{\NoHyper\footnotetext{These authors contributed equally to this work.}\endNoHyper}
\def\thefootnote{\arabic{footnote}}
\maketitle
\pagestyle{empty}

\begin{abstract}
A battery management system (BMS) relies on real-time estimation of battery temperature distribution in battery cells to ensure safe and optimal operation of Lithium-ion batteries. However, physical BMS often suffers from memory and computational resource limitations required by high-fidelity models. Temperature estimation of batteries for safety-critical systems using physics-based models on physical BMS can potentially become challenging due to their higher computational time. In contrast, neural network-based approaches offer faster estimation but require greater memory overhead. To address these challenges, we propose Kolmogorov-Arnold network (KAN) based thermal model, KAN-therm, to estimate the core temperature of a cylindrical battery. Unlike traditional neural network architectures, KAN uses learnable nonlinear activation functions that can effectively capture system complexity using relatively lean models.   We have compared the memory overhead and estimation time of our model with state-of-the-art neural network and tree-based models to demonstrate the applicability and potential scalability of KAN-therm on a physical BMS.
\end{abstract}

\section{Introduction}
Reliable operation of Lithium-ion batteries plays a crucial role in ensuring the efficient operation of renewable energy-based infrastructure systems, including smart grids and electric vehicles \cite{bukhari2024renewable}. This reliability is heavily dependent on the operating temperature of the batteries, and effective thermal management can ensure optimal battery operation, improve longevity, and prevent overheating \cite{karnehm2025core}. However, unlike the surface temperature, the battery core temperature is often inaccessible; thus, {estimating the temperature distribution within a battery cell} is critical to ensure safety \cite{chen2020core}. While high-fidelity thermal model-based $H_\infty$ observers \cite{lin2019robust} and Kalman-filters \cite{chen2020core} have been adopted for core temperature estimation, these methods are not well-suited for practical application due to their high computational demand, parameterization difficulty, and poor observability issues \cite{zheng2024real,ghosh2025detection}. Therefore, recent battery research has focused on developing advanced machine learning algorithms to capture the thermal behavior of batteries \cite{karnehm2025core}.

\subsection{Data-driven battery thermal models}
Data-driven battery thermal models estimate various thermal states and characteristics of the battery from measured signals, such as current, voltage, and ambient conditions. For instance, \cite{zhang2024benchmarking} developed three high-performing machine learning models for core temperature estimation based on a recurrent neural network (RNN), a long short-term memory (LSTM) network, and a gated recurrent unit (GRU) network. 
Similarly, \cite{zheng2024real} proposed an LSTM-RNN-based model to generate a real-time estimation of the average battery temperature. Accordingly, \cite{zhu2021data} developed an LSTM-based model for estimating instantaneous thermal states as well as for forecasting long-term thermal fluctuations to capture any early warning signs of thermal anomalies. \cite{ouyang2023data} deployed separate LSTM models for each cell in a battery to generate parallel cell temperature estimation for reliable thermal runaway estimation. On the other hand, \cite{hasan2020data} proposed a bank of seasonal nonlinear autoregressive exogenous neural network models corresponding to the weather characteristics in each season for cell temperature estimation. Furthermore, \cite{wang2021core} integrated LSTM model with transfer learning for temperature estimation of lithium-ion battery while reducing data requirement and facilitating adaptability across different batteries. Among recent research works, 
\cite{haraz2025hybrid} introduced an extra trees regressor-random forest (ETR-RF) model and a decision tree-Light Gradient Boosting Machine (DT-LGBM) for estimating the core temperature of electric vehicle batteries and showed 
that the latter exhibited superior computational speed in real-time applications. 
\cite{han2025transient} developed an LSTM-based nondimensionalized thermal model for transient core-temperature prediction under varying operating conditions. Furthermore, \cite{wang2025lithium} developed a Convolutional Neural Network–Bidirectional Long Short-Term Memory–Attention Mechanism (CNN–BiLSTM–AM) framework for predicting the internal temperature field of lithium-ion batteries under complex operating environments. This work highlights the increasing adoption of spatiotemporal feature learning architectures in data-driven battery thermal modeling.

\subsection{Hybrid battery thermal models}
Several works have also combined data-driven methods with physics-based models \cite{surya2022hybrid} to improve accuracy and interpretability. An unscented Kalman filter has been integrated with the neural network model to generate accurate temperature estimation with experimental validation in \cite{feng2020co}. Similarly, \cite{liu2022online} proposed a noise-compensated extended Kalman filter for core temperature estimation, where they utilized a neural network model to characterize the model noises from available sensor data.  In \cite{surya2022hybrid},  a two-dimensional grid LSTM model was adopted, where the model-based heat generation estimation was incorporated as one of the input features to improve the core temperature estimation performance further. Machine learning methods such as artificial neural network \cite{arora2017neural} and convolutional neural network combined with the artificial bee colony optimization technique \cite{yalccin2022cnn} have also been adopted for heat generation rate estimation in battery cells. Furthermore, to obtain reduced model complexity in the hybrid models, \cite{yuan2024core} utilized numerical techniques to simplify the thermal model, and then fused the numerical model with an LSTM network for core temperature estimation.   \cite{jia2025online} coupled an aging integrated electro thermal equivalent circuit model with a joint extended Kalman filtering framework for simultaneous estimation of battery core temperature, state-of-charge, and capacity. Similarly, \cite{shen2025physics} incorporated battery thermal physics into a recurrent learning framework to improve robustness under varying operating conditions. Furthermore, \cite{yuan2025online} proposed a physics-fused backpropagation neural network for online heat-generation and temperature prediction, and also demonstrated the effectiveness of coupling simplified thermal physics with neural estimators.  

\subsection{Research gap}
Model-based estimation methods solve the battery's thermal equations using numerical integration, which requires considerable computational resources. The required computations increase estimation latency, which can potentially limit their applicability in safety-critical applications such as the detection of short circuits \cite{ghosh2025koopman}. On the other hand, all neural network models proposed in the literature require a significantly large memory overhead to store their network structure. More sophisticated models require more memory, and with that, the time for forward computation also increases. These models can be too large for BMS hardware due to constraints in terms of the availability of processing power and memory capacity, creating a dichotomy between the desired accuracy and real-world practicality.  For instance, the FRAM of foxBMS\textsuperscript\textregistered\,  is 2 Mbit. In this context, a battery digital twin can be used to perform the computationally intensive calculations in the cloud BMS to support the computational demands of high-fidelity multi-physics models \cite{kim2018cloud,goldsworthy2022cloud} and provide desired accuracy without overloading the local BMS controller. However, cloud-based BMS face challenges due to data privacy concerns, communication delay, and poor connectivity. Moreover, small BMS are also essential for applications such as health monitoring devices and embedded electronics \cite{chen2023bms}. Additionally, the lightweight model developed for cells can be readily scaled up to the whole battery pack. Simple models run faster, which is important for quick safety monitoring and fault detection. This work addresses these research gaps by developing a lightweight data-driven KAN architecture for thermal modeling of batteries that is both accurate and efficient for safety-critical applications.\\

\subsection{Why KAN?}
Traditional neural networks, such as MLPs, learn the nonlinear relationships in the data by combining learnable linear weights at the edges with fixed nonlinear activation functions at the output nodes.
Under suitable conditions, they are known to be universal approximators because they can represent a wide range of nonlinear functions \cite{cybenko1989approximation}. For sequential data, recurrent neural networks are commonly used. LSTMs extend standard recurrent models by introducing gated memory to better handle long-term dependencies and training difficulties such as vanishing gradients \cite{luo2025physics}.  In contrast, KANs replace scalar weights with learnable univariate functions on edges, where these functions are spline-parameterized instead of fixed activations \cite{liu2024kan}. This design is motivated by the Kolmogorov–Arnold representation theorem, which expresses multivariate functions as combinations of univariate functions. Tree-based models such as eXtreme Gradient Boosting (XGBoost) are also used for regression-based estimation \cite{chen2016xgboost}. Similar to neural networks, these methods require a large number of trees for acceptable accuracy, which increases memory overhead and inference time. While classification models using XGBoost can be easily interpreted, in case of regression tasks this interpretability is substantially lost due to the large size of the model and continuous nature of regression-based estimation. 

Overall, KANs offer three main advantages. First, they can be more parameter-efficient, \cite{liu2024kan} shows that smaller KANs can match or outperform larger MLPs in function-fitting and PDE-related tasks. Second, they are more interpretable, because each connection is a learnable univariate function that can be directly visualized, making it easier to understand how inputs affect the output. 
Third, KANs also provide analytical guarantees on error. If the target function is smooth, the approximation error reduces in a predictable way as the model resolution increases, with the rate depending on the function's smoothness and spline order. This helps to clearly relate the model design to its accuracy, instead of relying only on experiments and post-analysis.

\subsection{KAN battery models}
KAN is becoming increasingly popular in the battery community due to its lean structure, accuracy, and interpretability \cite{liu2024kan,yang2025capacity}. For example, \cite{sulaiman2024battery} has shown that the KAN model outperforms artificial neural network and barnacles mating optimizer-deep learning models in battery state-of-charge (SOC) estimation.  Moreover,  hybrid convolutional neural network–KAN model \cite{cui2025enhanced},  Kolmogorov–Arnold–Linformer network model \cite{shao2025soh}; and KAN model integrated with dynamic-graph-generator \cite{liu2025soh} have been explored for predicting battery state-of-health with improved accuracy while utilizing the learnable activation functions of KAN to enhance the model's robustness, flexibility, and the capability for nonlinear features representation. Similarly, \cite{he2025remaining}  combined KAN with a lightweight GRU network to enhance the generalization and feature extraction capability of the GRU, leading to an improved performance in the remaining useful life estimation for batteries.  \cite{yang2025capacity} combined KAN with a squeeze-and-excitation module for predicting the capacity degradation of electric vehicle battery packs, where the KAN module improved the model’s adaptability to changes in charging data. 
  \cite{liu2025physical} introduced a physical information-guided KAN for lithium-ion battery SOH estimation. Moreover, \cite{jarraya2025soh} proposed hybrid KAN-LSTM models for enhanced SOH monitoring of Li-ion batteries. CNN–Temporal-KAN architectures have also been adopted in \cite{liu2025state} for SOC estimation to improve feature extraction and sequential estimation performance. \cite{wang2025synergistic} proposed a hybrid RNN-KAN framework with optimized feature extraction for lithium-ion battery SOH estimation. In addition, an attention-enhanced physics-informed KAN model was developed in \cite{wei2025predicting} for lithium-ion battery SOH estimation.
 
The success of KAN for battery state estimation motivates our research. Additionally, \cite{karnehm2025core} focused on estimating the core temperature of batteries without surface-temperature feedback, but at the cost of a complex network architecture with higher computational demand. This work highlighted the need for the refinement of the KAN architecture for core temperature estimation that is geared towards implementation on a physical BMS.

\subsection{Contributions}
Our primary \textit{contribution} in this work is to develop a data-driven thermal model for cylindrical cells to obtain reliable real-time core temperature estimation.  This model is developed using the KAN architecture as it learns nonlinear input-temperature relationships through learnable activation functions.  Unlike other neural network models, the flexibility in learning nonlinear activations reduces the need for learning extensive linear combinations with fixed nonlinear activation to approximate nonlinear relationships among inputs and outputs. Through an exhaustive and tailored hyperparameter grid search, we identified the optimal functional regularization penalties, activation function choices, and update controls that \textit{push the limits of achievable network compactness and efficiency while preserving practicable accuracy}. To train this model, we also identified an optimal selection of (features or) battery measurements that can provide both the desired accuracy and model simplicity. MLP-, LSTM-, and RNN-based models have been used here as baselines for comparison, since they are the most common and reliable high-accuracy network structures used in data-driven temperature estimation and general sequential learning. 

The rest of the paper is organized as follows. Section~\ref{prel} presents the preliminaries and theoretical background for KAN. The proposed KAN-therm model development and training are described in Section~\ref{KANtherm}. We present the detailed results of our performance comparison in Section~\ref{sim}. Finally, in Section~\ref{conclusion}, we conclude our work.







\section{Preliminaries On Kolomogorov-Arnold Network} \label{prel}
The Kolmogorov-Arnold network (KAN) \cite{liu2024kan} is built upon the Kolmogorov-Arnold representation theorem, which states that a multivariate continuous function can be expressed as a finite composition of univariate continuous functions and the addition operation. While the theorem asserts the existence of such univariate functions, their construction is not straightforward. KAN utilizes an $L$-deep multi-layer network structure to learn these functions, where the $l^{th}$ layer has $W_l$ number of nodes and each node represents a single variable $x_{l,w},\, \forall w\in\{1,\hdots, W_l\},\, l\in \{1, \hdots, L\}$. $W_l$ here represents the width of layer $l$. At each KAN-layer $l$, the value $x_{l,w}$ at every node is passed through an activation function $\theta_{l,w^+,w},\, \forall w^+\in \{1, \hdots, W_{l+1}\}$ which are then added to produce the node value at each node $w^+$ in the consecutive layer $l+1$. This can be written as
\begin{align}
    x_{l+1,w^+}=\sum_{w=1}^{W_l}\theta_{l,w^+,w} (x_{l,w}), \, \forall w^+\in \{1, \hdots, W_{l+1}\}.
\end{align}
 \noindent In a compact matrix form, it can be expressed as, 
\begin{equation} \label{KAN matrix}
    x_{l+1} = \underbrace{ \begin{pmatrix}
\theta_{l,1,1}(\cdot) & \dots & \theta_{l,1,W_{l}}(\cdot)\\
\vdots & \ddots & \vdots\\
\theta_{l,W_{l+1},1}(\cdot) & \dots & \theta_{l,W_{l+1},W_{l}}(\cdot)
    \end{pmatrix}}_{\Theta_{l}}x_{l},
\end{equation}
where the activation function matrix at layer $l$ is given by $\Theta_l$ and the node vector at layer $l$ is denoted by $x_l, \, \forall l.$ The goal of a multi-layer KAN architecture is to learn these activation functions $\Theta_l$ $\forall l$ to capture the relationship between $N_1$ input variables (or feature variables) denoted by $x_1=[x_{1,1}, \hdots, x_{1, N_1}]$ and  $N_{L+1}$ output variables (or target/estimation variables) denoted by $x_o$. Mathematically, this can be represented as  $x_{o}=KAN(x_1)=(\Theta_L\circ \Theta_{L-1}\circ \hdots \circ \Theta_1)x_1$. In this work, we use a sum of silu function and a linear combination of $k$-order B-splines over $G$ grid points to represent our activation functions $\theta_{l,w^+,w}$. The weights for the spline combination and the ``fine-graining" of the grid are learned using backpropagation and gradient descent using the loss function defined below.

\textit{Loss function:} The loss function $\ell_{total}$ \eqref{estimation loss} used for KAN training is defined to reduce estimation error and the number of parameters of the KAN model, such that a lightweight structure is realized by prioritizing dominant activation functions \cite{liu2024kan}. The first objective is captured by the mean squared estimation error $\ell_{pred}$ between the predicted and true outputs. The sparsification of the KAN model through the reduction of its parameter count is realized by minimizing the $l_1$-norm contribution of the activation function, which is denoted by $|\Theta_l|_1$ in \eqref{l1}, and minimizing a self-entropy term $S(\Theta_l)$, defined in \eqref{entropy}. 
\begin{align} 
\ell_{total} &= \ell_{pred} + \lambda\Bigg(\nu_1\sum^{L-1}_{l=0}|\Theta_l|_1+\nu_2\sum^{L-1}_{l=0}S(\Theta_l)\Bigg), \label{estimation loss} \\
\left|\Theta_l\right|_{1} &= \sum_{w=1}^{W_l} \sum_{w^+=1}^{W_{l+1}} \left|\theta_{w,w^+}\right|_1, \label{l1} \\
S(\Theta_l) &= - \sum_{w=1}^{W_l} \sum_{w^+=1}^{W_{l+1}} \frac{|\theta_{w,w^+}|_1}{|\Theta_l|_1}\text{log}\bigg( \frac{|\theta_{w,w^+}|_1}{|\Theta_l|_1} \bigg), \label{entropy}
\end{align}
Here, $\lambda$ is the sparsification penalty term and $\nu_1,\nu_2$ are regularization hyperparameters of the model.

\textit{Parameter count:} The total number of parameters in the KAN network is given by $\mathcal{O}(\sum_{l=1}^{L-1}W_lW_{l+1}(G+k))$, for a network having depth $L$ and each spline is of order $k$ over $G$ intervals or for $(G+1)$ grid points.  

We note here that a standard neural network architecture, in contrast, is based on approximating functions of multiple variables (or input features) as \textit{learnable} linear transformations of node values through \textit{weights} and their subsequent transformation via \textit{fixed nonlinear activation functions}.

\section{Introduction to KAN-Therm Model} \label{KANtherm}
In this section, we develop our KAN-therm model to estimate the core temperature $\widehat{T}_1$ of the battery. We also discuss the data-generation strategy for training, validation, and testing the model using a battery-lumped thermal model. The details of the training procedure and hyperparameter selection have also been presented in this section.

\subsection{Battery thermal model} \label{therm}
To generate the battery data for this work, we adopt a lumped parameter thermal model of a cylindrical battery cell with a surface cooling system  \cite{lin2014lumped, vyas2022thermal}. The governing equations for the core temperature $T_1$, surface temperature $T_2$ of the cell, and coolant temperature $T_\infty$ of the cooling system are given by,
\begin{align} \label{thermal dynamics}
    \dot{T_1}(t) & = -\frac{T_1(t)-T_2(t)}{R_1C_1} + \frac{\dot{Q}(t)}{C_1},\\
    \dot{T_2}(t) & = -\frac{T_2(t)-T_1(t)}{R_1C_2} - \frac{T_2(t)-T_{\infty}(t)}{R_2C_2},\\ 
    \dot T_{\infty}(t) & = -\frac{T_{\infty}(t)-T_2(t)}{R_2C_{\infty}} - \frac{\dot Q_c(t)}{C_{\infty}},
\end{align}

\noindent where $R_1$ and $R_2$ are the thermal resistances between battery core and surface and between surface and coolant, respectively, $C_1$, $C_2$ and $C_\infty$ are the heat capacities of the battery material at core, at surface and of cooling system, respectively, $\dot Q_c$ represents cooling power of the cooling system, and $\dot Q$ is the battery internal heat generation term.
$\dot{Q}$ is related to the electrical behavior of the battery as
\begin{align} \label{ecm heat}
    \dot{Q}(t) = I(t)\bigg(V_{\text{OCV}}(SOC)-V_t(t)-T_1(t)\mathbb{E}\bigg),
\end{align}
\noindent where $I$ is the current through the battery, $V_{\text{OCV}}$ is the open circuit voltage as a function of SOC, $V_t$ is the terminal voltage, and $\mathbb{E}$ is the entropic heat coefficient of the battery cell.
\noindent The  SOC dynamics and terminal voltage $V_t$ of the battery is given by,
\begin{align} \label{electrical dynamics}
   \dot{SOC}(t) = -\frac{I(t)}{Q_b}, \quad
    V_t(t) = V_{\text{OCV}}(SOC) - I(t)R_s. 
\end{align}
\noindent Here, $Q_b$ is the battery capacity and $R_s$ is the battery internal resistance \cite{bernardi1985general}. Using \eqref{electrical dynamics} in \eqref{ecm heat}, we can represent $\dot{Q} = R_sI^2(t)-\mathbb{E}I(t)T_1(t) $.

\subsection{Data generation}
\begin{figure}[ht]
    \centering
    \includegraphics[width=1\linewidth]{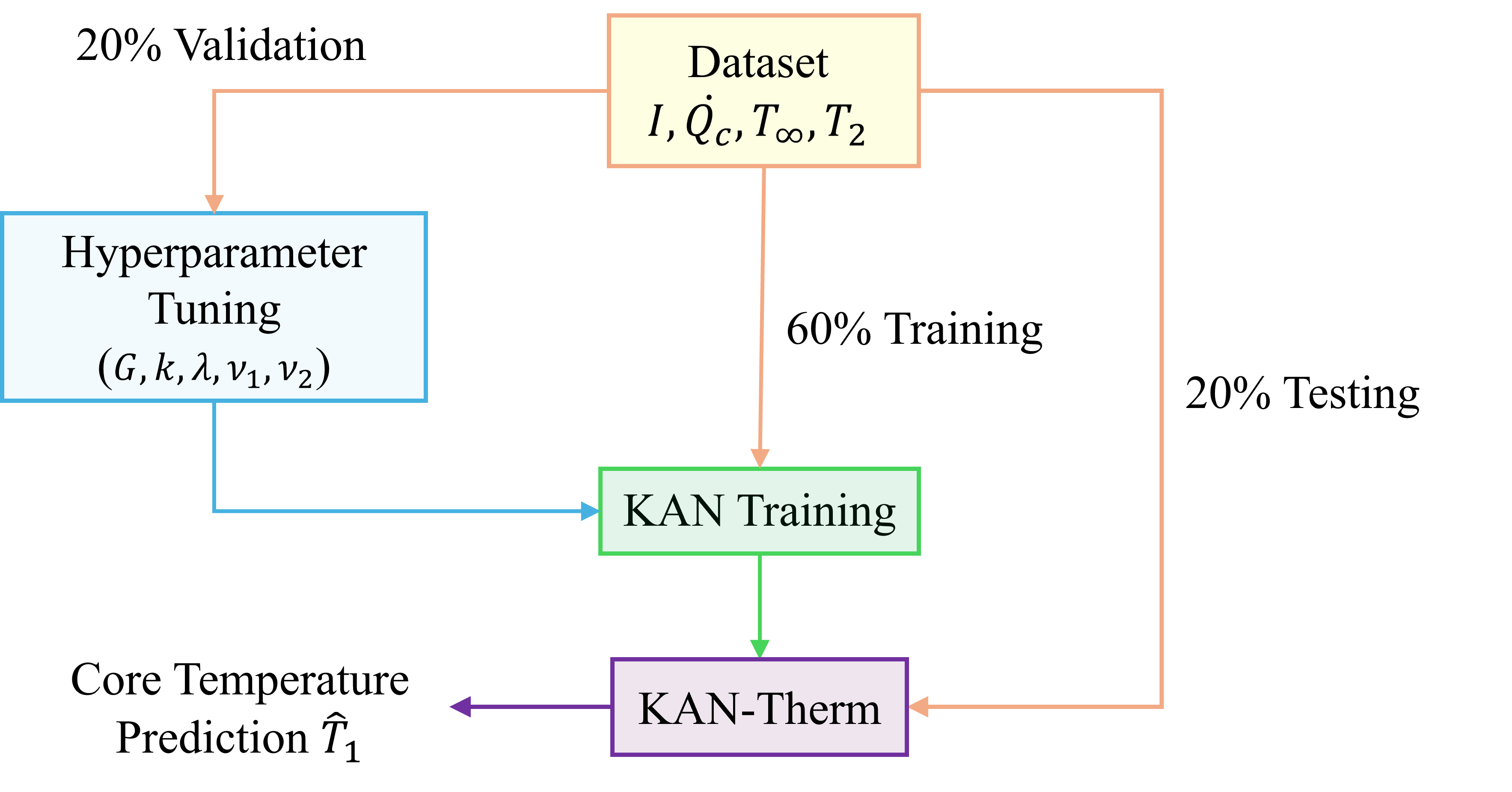}
    \caption{Flowchart illustrating how the dataset is split into training, validation, and testing subsets.}
    \label{fig:flow}
\end{figure}

\begin{figure}[ht]
    \centering
    \includegraphics[width=1\linewidth]{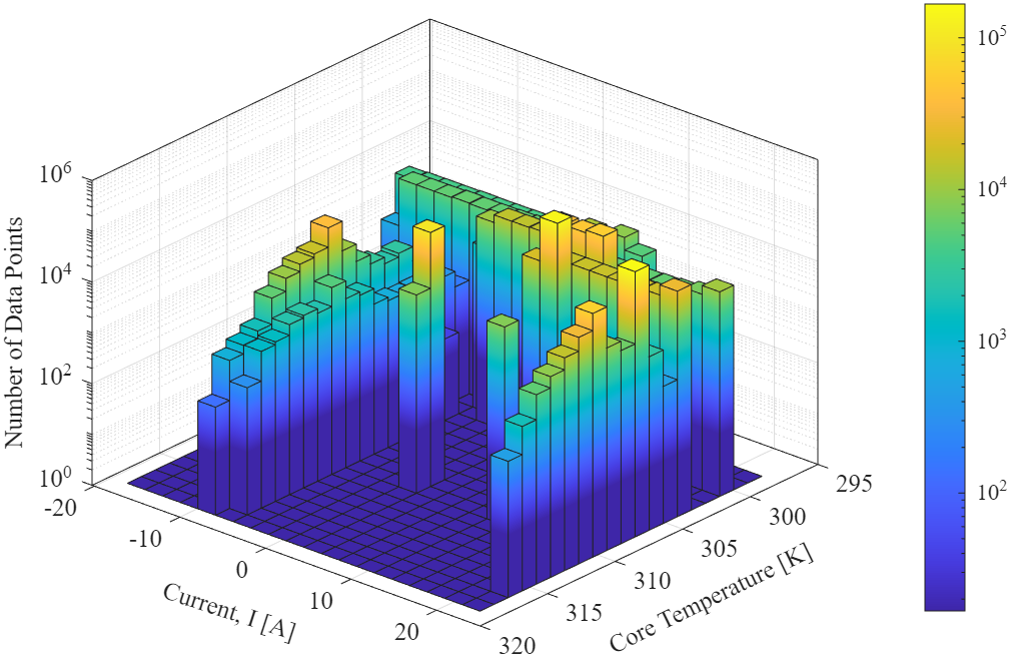}
    \caption{Histogram showing training data distribution of current $I$ vs core temperature $T_1$.}
    \label{fig:histogram}
\end{figure}

For this study, we consider a $2.3Ah$ 
cylindrical 
$LiFePO_4-LiC_6$ battery cell where the parameters for \eqref{thermal dynamics}-\eqref{electrical dynamics} are adopted from \cite{feng2024safe, dey2019safer} and are presented in Table~\ref{tab:PARAMETERS}.
We use numerical integration in Python for the lumped-parameter thermal model \eqref{thermal dynamics}-\eqref{electrical dynamics} to collect training, validation, and test data under different charging-discharging scenarios. The dynamic current profiles were selected from standard automotive drive cycles: the Urban Dynamometer Driving Schedule (UDDS) \cite{kruse1973development} for training, the Supplemental Federal Test Procedure (US06) \cite{ahmadzadeh2023physics} for validation, and the VEIL2NREL profile for testing from the automotive battery cycle dataset \cite{luzi2018automotive}. The constant current (CC) profiles varied from $0.9$C-rate ($2.07$A) to $10$C-rate ($23$A). Moreover, the dataset accounts for variations in coolant power $\dot{Q}_c$, initial SOC, and initial battery temperature (i.e., ambient temperature). In total, time series datasets from $18$ distinct scenarios have been obtained, and $60\%$ of this data has been used to train the KAN-therm model, $20\%$ data to validate, and $20\%$ data for testing the performance of the proposed KAN-therm model, as shown in Fig.~\ref{fig:flow}.  The testing dataset also includes data from constant-current charging at a 3C rate. Furthermore, to simulate realistic sensor measurements,  zero mean Gaussian noise with a standard deviation of $0.015$K has been added to the temperature measurements and a standard deviation of $0.006 A$ for the current signal. The histogram in Fig.~\ref{fig:histogram} shows the distribution of training data across different ranges of current and core temperature for the battery.


\begin{table}[ht]
    \centering
    \renewcommand{\arraystretch}{1.4} 
    \setlength{\tabcolsep}{8pt} 
    \begin{tabular}{|c|c||c|c|}
    \hline
    \textbf{Parameters} & \textbf{Values} &  \textbf{Parameters} & \textbf{Values}  \\
    \hline
    $R_s$ & $1.00$ $\mathrm{m}\Omega$ &$Q_b$ & 2.3 Ah \\
    \hline
    $R_1$ & $1.94$ $KW^{-1}$  & $R_2$ & $3.08$ $KW^{-1}$ \\
    \hline
    $C_1$ & $62.70$ $JK^{-1}$ &$C_2$ & $4.50$ $JK^{-1}$  \\
    \hline
    $C_\infty$ & $10.00$ $JK^{-1}$ & $\mathbb{E}$ & $10^{-4}$ $WK^{-1}$\\
    \hline
    \end{tabular}
    \vspace{1mm}
    \caption{Parameters used for battery data generation.}
    \label{tab:PARAMETERS}
\end{table}

\subsection{Model development}
\begin{figure*}[ht]
    \centering
    \includegraphics[width=1\linewidth]{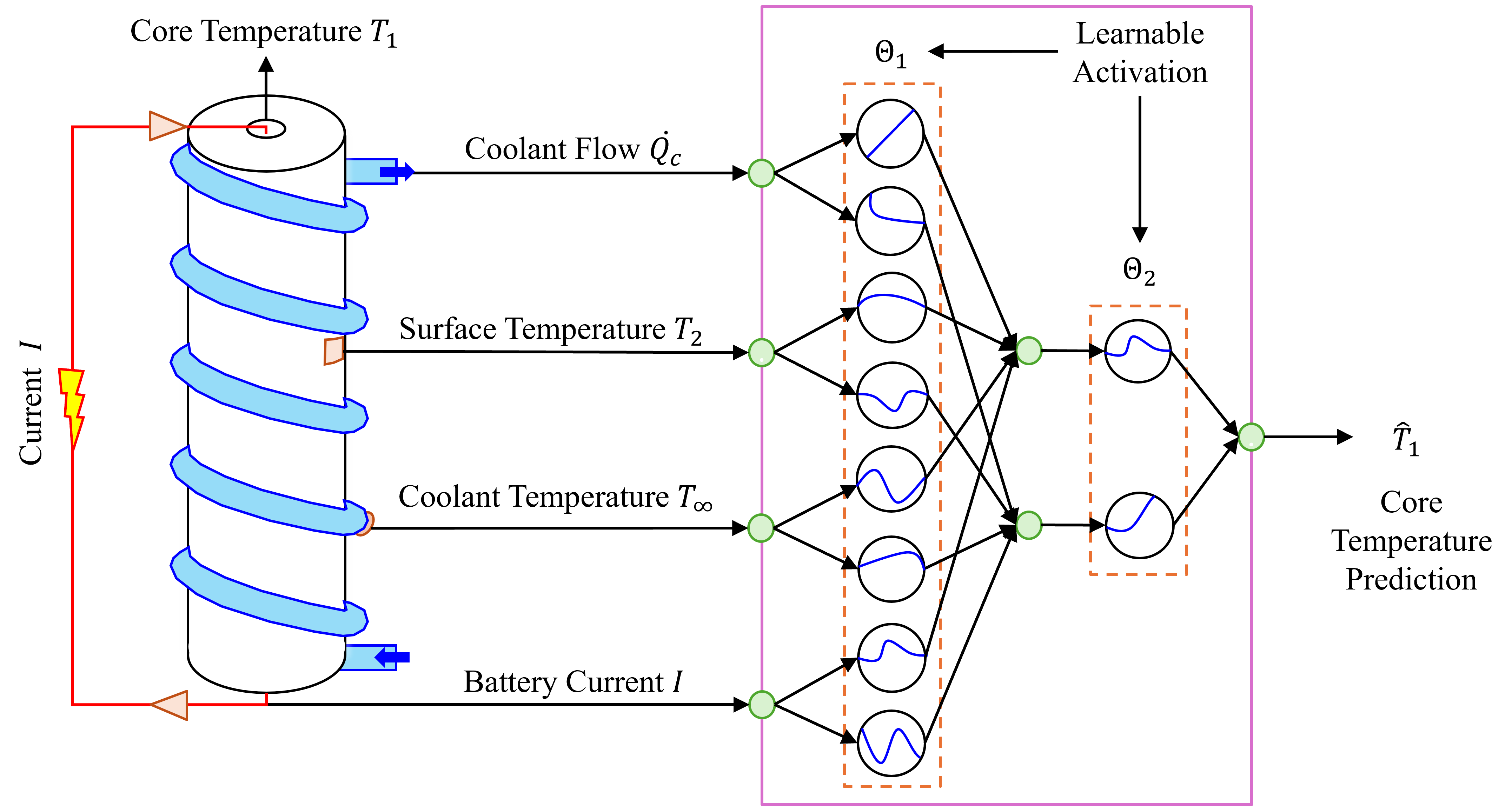}
    \caption{Block diagram of the proposed KAN-therm model for estimation of battery core temperature $\widehat{T}_1$.}
    \label{fig:blockdiag}
\end{figure*}

The proposed KAN-therm model incorporates $4$ input features: the applied current $I$, the cooling power $\dot{Q_c}$, the temperature of the cooling system $T_\infty$, and the surface temperature of the battery $T_2$. These variables can be readily measured and are assumed to be known. Our goal in training the KAN-therm model is to learn the activation functions such that $\widehat{T}_1 = \text{KAN}(I,\dot{Q}_c,T_{\infty},T_2)$, based on the training data generated (as described in the last subsection). Fig.~\ref{fig:blockdiag} shows the block diagram of the proposed KAN-therm model, with its input features and output. 

For our KAN-therm model, the four input variables $I,\dot{Q}_c,T_{\infty}, $ and $ T_2$ are input nodes of the $1^{st}$ layer of width $N_1 = 4$. Next, we consider a KAN layer of width $N_2 = 1$, to finally produce the core temperature estimation output $\widehat{T}_1$. Thus,  we have a network configuration of $[\![4, 1, 1 ]\!]$ for the KAN-therm model. In this architecture, $4$ learnable activation functions are present in the $1^{st}$ KAN layer, while $1$ is present in the $2^{nd}$ KAN layer. To train the KAN-therm model, the complete dataset has been normalized between $0$ to $1$ range using min-max normalization. Furthermore, to efficiently optimize the minimization of the loss function, L-BFGS optimizer has been used during training. 

\noindent
 \textbf{Hyperparameter tuning:}
 We adopt a \textit{ grid-based search} approach to obtain the best-suited set of hyperparameters that identifies the KAN-therm model that minimizes the loss function $\ell_{total}$  \eqref{estimation loss} on the validation dataset. Grid search is a brute-force or exhaustive way of searching through a manually specified subset of the hyperparameter space, defined by lower and upper bounds and specific steps \cite{claesen2014easy}. During hyperparameter tuning, we grid-searched nine hyperparameters of the KAN-Therm model, as described below.  

   \textit{\uline{Network width}} for KAN layers affects the model complexity as well as the learning capability. As part of our hyperparameter search,  the width of the $2^{nd}$ KAN layer ($l =1$) is varied over the values $\{1, 2, 3, 4\}$. We also tested KAN structures with more than one layer, however they produced comparable or lower accuracy, with a significant increase in parameter count. We have not presented the analysis of those structures in this paper for brevity.
 
 \textit{\uline{Spline order $k$ \& number of grid intervals $G$}} significantly influence the nonlinear feature representation capability of the model. For both of these hyperparameters, the model performance is tested over the set  $\{2, 3, 4,5,6 \}$. The grid search results indicate that $k=3$ and $G=2$ yield the best model performance.
 
\textit{\uline{Sparsification parameter $\lambda$ and regularization parameters $\nu_1, \nu_2$}} ensure parameter reduction of the model by suppressing weaker activations. During grid search, we consider the set $\{0.00001, 0.00005, 0.0001 \}$ for $\lambda$ and the set $\{0.14, 0.15, 0.17, 0.20\}$ for both $\nu_1, \nu_2$. Based on our grid search results, we choose $\lambda = 0.00005, \nu_1=0.14, \nu_2 = 0.17$ to obtain the best accuracy, while maintaining the lightweight structure for the model.
 
 \textit{\uline{Total number of epochs \&  early stopping of the spline grid update}} impact the model convergence and the chances of overfitting. During our model training, we monitor the convergence of training and the validation loss over a set of epoch choices $\{60,70, 50, 90\}$ to obtain the best model performance for $70$ epochs. Similarly, the spline grid update hyperparameter controls the number of epochs over which the spline grid range is updated to minimize the loss function. 
During model training, we consider a set of spline-grid stopping fractions $\{0.70,0.80, 0.90\}$ of the total epochs and select the best configuration based on the validation results to avoid overfitting. For our best model, we have the training run of $70$ epochs with a grid-update stopping fraction of $0.9$ i.e. the spline-grid updates were stopped after 63 epochs.
 
\uline{\textit{Training batch size}} affects the convergence, computational speed, and resource requirement during training.  The batch size during our model training is finalized as $2^{14}$ after searching through the set of batch size choices of $\{2^{14}, 2^{15}, 2^{16}\}$. 

\begin{figure*}[ht]
      \centering 
    \includegraphics[width=1\linewidth, trim = {0cm 0cm 0.45cm 0cm},clip]{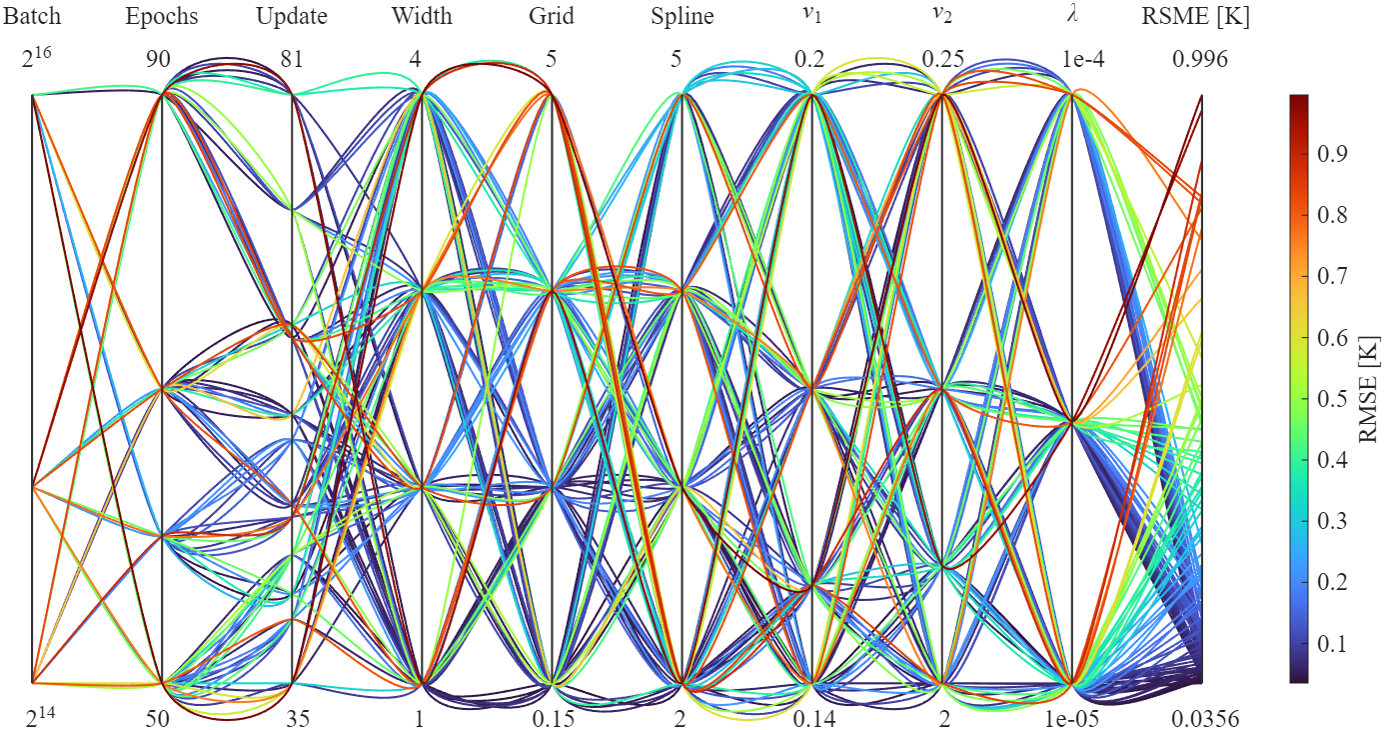} 
    \caption{Parallel plot showing the validation errors of KAN-Therm for a grid search over nine hyperparameters, namely batch size, number of steps, early stopping of the grid update, network width, grid interval $G$, spline order $k$, sparsification parameter $\lambda$, and regularization parameters $\nu_1$ and $\nu_2$.}
    \label{fig:hyper}
\end{figure*}

Fig.~\ref{fig:hyper} summarizes the KAN sweep results, where each line corresponds to a trained model and shows the settings (batch size, epochs, update steps, width, grid, spline, $v_1$, $v_2$, $\lambda$) used in that run. The color scale directly shows RMSE (K), with lower RMSE indicating better models. Table~\ref{tab:MODEL_VARIABLES_AND_PARAMETERS} shows the specific hyperparameter settings for the chosen best-performing KAN model (lowest RMSE in the sweep).

The parallel coordinate plot in Fig.~\ref{fig:hyper} shows the results of the hyperparameter search, where each vertical axis represents a hyperparameter and each colored line connects the hyperparameter combinations chosen for each training-validation run of the model. The color of each line maps to the validation RMSE for that run. The hyperparameter combination that yields the best model performance with  1.83E$-6$ training loss and 1.96E$-6$ validation loss is listed in Table~\ref{tab:MODEL_VARIABLES_AND_PARAMETERS}. Next, we test the performance of our proposed KAN-therm model and highlight its advantages over state-of-the-art machine learning networks, specifically in terms of deployability on a resource-constrained physical BMS. 

\begin{table}[ht]
    \centering
    \renewcommand{\arraystretch}{1.4} 
    \setlength{\tabcolsep}{8pt} 
    \begin{tabular}{|c|c||c|c|}
    \hline
\textbf{Hyperparameters} & \textbf{Values} & \textbf{Hyperarameters} & \textbf{Values} \\   
    \hline
    Network structure & $[\![4,1,1 ]\!]$ & $G$ & $2$ \\
    \hline
   Epochs & $70$ & $k$ & $3$\\
   \hline
    Batch size & $2^{14}$ & $\lambda$ & $0.00005$   \\
    \hline
     $\nu_1$ & $0.14$ & $\nu_2$ & $0.17$\\
    \hline
    \end{tabular}
    \vspace{1mm}
    \caption{Details of the final KAN-therm model.}
    \label{tab:MODEL_VARIABLES_AND_PARAMETERS}
\end{table}

\section{Performance Tests And Comparison} \label{sim}
In this section, we present test results comparing the performance of the proposed KAN-therm model with four benchmark machine learning networks: MLP, RNN, LSTM, and XGBoost. This comparison allowed us to provide a well-rounded assessment of the KAN-therm model's predictive performance, memory requirements, and computation time.
\subsection{Baseline methods: MLP, RNN, LSTM \& {XGBoost}}
We compare the proposed KAN-therm model with baseline three neural network models and a {tree-based ensemble model} across five performance criteria: number of parameters, estimation time, training RMSE, validation RMSE, and testing RMSE.
We thoroughly explored the hyperparameter space for each baseline method via extensive grid search and compared the best-performing configurations with our proposed KAN-therm model. Furthermore, we used the ReLU activation function and the Adam optimization method in all {neural network} models to mitigate the risk of vanishing gradients, allowing faster training convergence \cite{tan2019vanishing}. 

\textbf{MLP} is a type of feedforward neural network consisting fully-connected layers \cite{shao2025soh}.
 Our baseline MLP is chosen as $[\![4,16,16,1 ]\!]$ i.e., there are $4$ neurons in the input layer, $16$ neurons in each of the two hidden layers, and a single output neuron. 

\textbf{RNN} is tailored for sequential data processing that contains neural networks with recurrent units, a form of memory that is updated at each time step based on the current input and the hidden state from the previous step \cite{zhang2024benchmarking}. In this study, we consider the baseline RNN model with $[\![4, 8,1]\!]$ structure, representing a recurrent layer with $8$ memory units and a fully connected output layer. This model is trained using a lookback window of $15$ data points for $62$ epochs using a batch size of $2048$ and a learning rate of $0.0005$.  

\textbf{LSTM} is a variant of RNN that can address the issue of long-term dependency and learns to store and discard information selectively for efficient learning \cite{zhang2024benchmarking}.
LSTM includes one cell and three gates: forget, input, and output gate such that the cell state holds the values for arbitrary time interval. The forget gate removes information that is no longer needed in the cell state, the input gate adds useful information  in the cell state, and the output gate extracts this useful information 
as the output \cite{zhang2024benchmarking}. Based on the hyperparameter search, we chose our baseline LSTM network structure as $[\![4,12,1 ]\!]$ with $12$ LSTM units in the intermediate layer. This model is trained with a lookback window of $15$ data points, learning rate of $0.0005$, batch size $4096$, and $128$ epochs.

{\textbf{XGBoost}} adopts the gradient tree booster strategy to sequentially train an ensemble of trees such that each tree assimilates the prediction error in the previous tree ensemble \cite{chen2016xgboost, ghosh2024koopman}.
Form our hyperparameter search, we choose the best model with 3039 trees where each tree can be split twice. We consider a learning rate of 0.01 with $l1$ and $l2$ regularization parameters of  0.01 and 0.0065, respectively, to train the model, while using  $70\%$ of the training data at each iteration step.
In addition, we provide the XGBoost model with an initial prediction for a given input as a base margin to obtain improved accuracy. We utilize a least-squares linear regression model to generate these base margin values.

\begin{figure}[ht]
    \centering
    \includegraphics[width=1.0\linewidth]{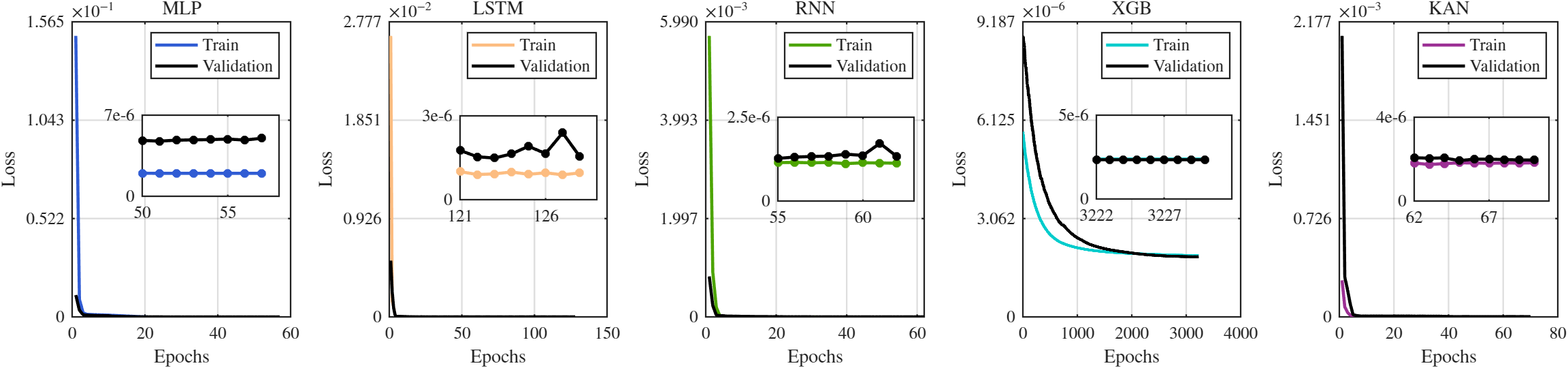}
    \caption{Plots showing the convergence of training and validation losses for MLP, LSTM, RNN, and KAN models.}
    \label{fig: loss}
\end{figure}

Fig.~\ref{fig: loss} shows the converging training and validation losses for each of the {five models}: MLP, LSTM, RNN, {XGBoost}, and KAN, indicating optimal model fitting. 

\subsection{Results \& comparison}
\begin{figure}[ht]
    \centering
    \includegraphics[width=0.9\linewidth]{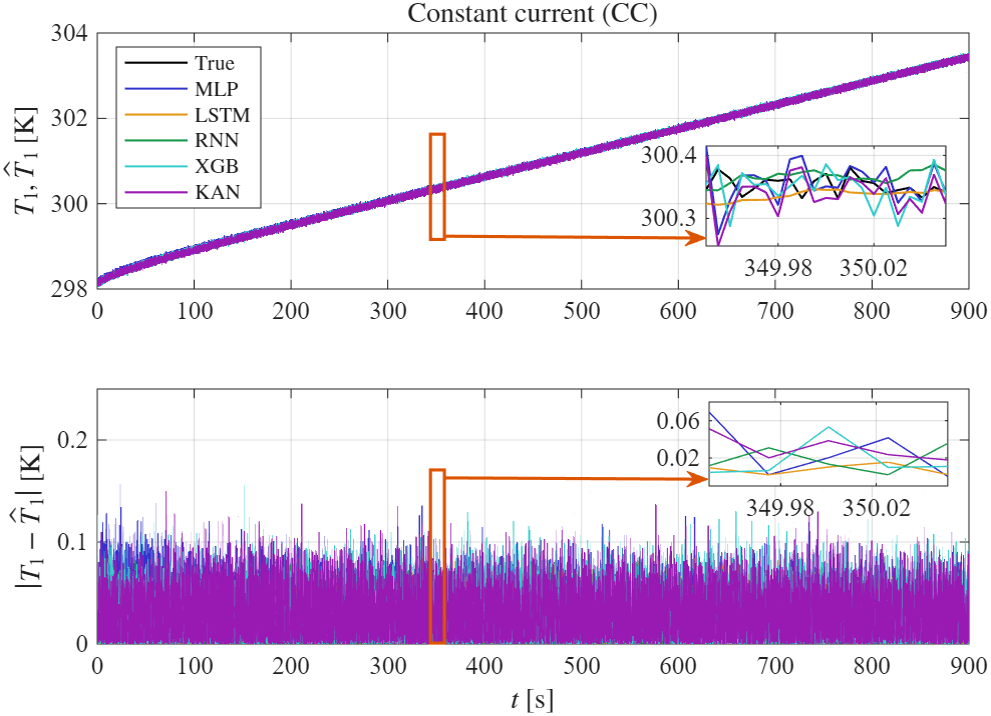}
    \caption{Under constant current charging test, the top plot shows the true core temperature $T_1$ and the predicted core temperature $\widehat{T}_1$ from four neural network models; the bottom plot presents the corresponding estimation error.}
    \label{fig:prediction_cc}
\end{figure}
\begin{figure}[ht]
    \centering
    \includegraphics[width=0.9\linewidth]{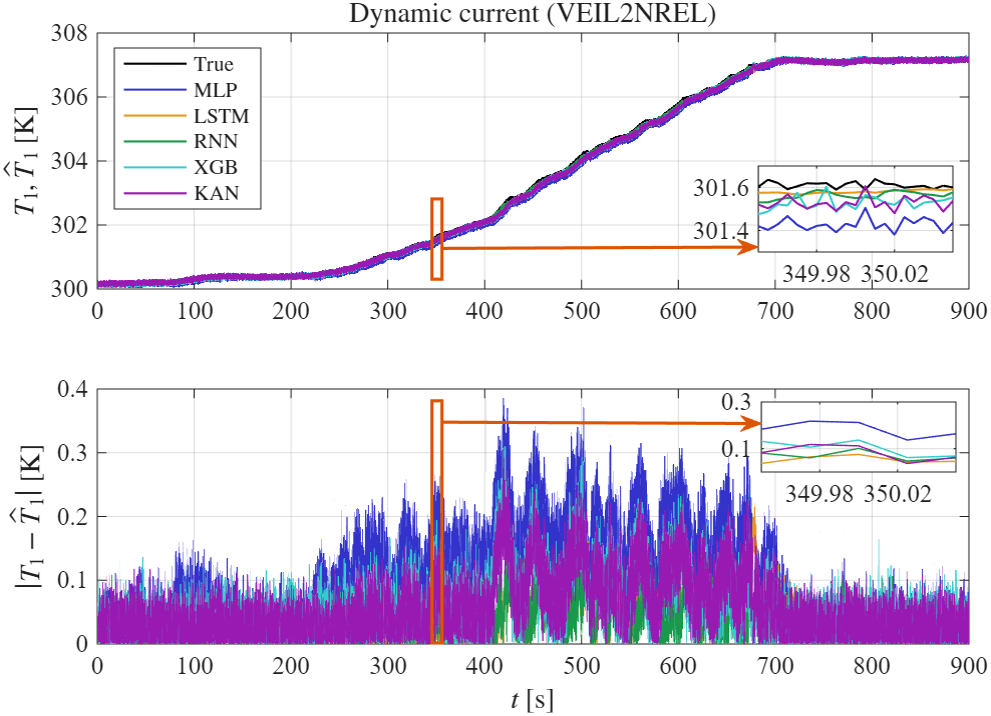}
    \caption{Under VEIL2NREL dynamic current test, top plot shows the true core temperature $T_1$ and the predicted core temperature $\widehat{T}_1$ from four neural network models; the bottom plot presents the corresponding estimation error.}
    \label{fig:prediction_udds}
\end{figure}
The models are evaluated using a test dataset containing a CC $3$C-rate charging data and the VEIL2NREL dynamic current profile from the Automotive Li-ion Cell Usage Data Set \cite{luzi2018automotive}. The latter contains high-frequency current fluctuations representing real-world driving conditions.

Fig.~\ref{fig:prediction_cc} and Fig.~\ref{fig:prediction_udds} illustrate the test results for the CC and VEIL2NREL dynamic current testing scenarios, respectively. The top plots of Fig.~\ref{fig:prediction_cc} and Fig.~\ref{fig:prediction_udds} show the true core temperature $T_1$ (obtained from the battery thermal model) along with the estimated core temperature $\widehat{T}_1$ using KAN-therm model and the three baseline models. The corresponding estimation errors (the absolute difference between true and predicted values) are shown in the bottom plots of Fig.~\ref{fig:prediction_cc} and Fig.~\ref{fig:prediction_udds}. 
Next, we present a detailed comparison of the performance of the KAN-therm model with the three baseline neural network models, {the tree-based ensemble model} and the physics-based model \eqref{thermal dynamics}-\eqref{electrical dynamics} across five criteria: the number of trainable parameters, estimation time, and  training, validation, and testing RMSE. Table~\ref{tab:per_compSumm} summarizes the results of this comparison.

\setlength\arrayrulewidth{1pt} 
\begin{table*}[!t]
    \centering
    \caption{Performance comparison of KAN-Therm vs different models.} 
    \renewcommand{\arraystretch}{2.2}
    \setlength{\tabcolsep}{6pt}
    \begin{NiceTabular}{|c|c|c|c|c|c|c|}
    \CodeBefore
    \cellcolor{blue!50!yellow!60!white}{2-2}%
    \cellcolor{blue!80!green!80!white}{2-3}
    \cellcolor{orange!50}{2-4}
    \cellcolor{green!50!black!70!yellow}{2-5}
    \cellcolor{cyan}{2-6}
  \cellcolor{blue!30!purple!80!white}{2-7} 
\Body
        \hline
        & \multicolumn{6}{c|}{\textbf{Models}} \\
        \cline{2-7}
        & \textbf{Physics} & \textbf{MLP} & \textbf{LSTM} & \textbf{RNN} & \textbf{XGBoost} & \textbf{KAN} \\
        \hline
        \textbf{Parameters} 
            & \fbox{9} & 369 & 877 & 121 & 6000 & \fbox{25} \\
        \hline
        \textbf{\shortstack{Estimation\\Time [$\mu$s]}} 
            & 1.3 & \fbox{0.0537} & 0.4945 & 0.8556 & 0.8050 & \fbox{0.0395} \\
        \hline
        \textbf{\shortstack{Train\\RMSE [K]}} 
            & NA & 0.0366 & 0.0251 & 0.0281 & 0.0358 & {0.0345} \\
        \hline
        \textbf{\shortstack{Validation\\RMSE [K]}} 
            & NA & 0.0581 & 0.0317 & 0.0288 & {0.0354} & {0.0358} \\
        \hline
        \textbf{\shortstack{Testing\\RMSE [K]}} 
            & NA & 0.0852 & 0.0359 & \fbox{0.0334} &  0.0534 & \fbox{0.0477} \\
        \hline
    \end{NiceTabular}
    \label{tab:per_compSumm}
\end{table*}

\textbf{Criterion 1: Parameters.} The models show large differences in the number of trainable parameters. The physics-based model is minimal, with only nine parameters. Among the learned models, KAN is the most compact with only $25$ learnable parameters. In comparison, the RNN is nearly five times larger, MLP and LSTM expand to approximately $15 \times$ and {$35\times$}, respectively. {Lastly, the  XGBoost model required approximately $240 \times$ more parameters compared to the KAN model.} This demonstrates the KAN-Therm's strength as a lightweight alternative for thermal modeling of a battery. \\

\textbf{Criterion 2: Estimation time.} We measure the time to estimate one sample of the test data for each method using Python’s \texttt{time.time()} function. For the neural network models, it is equal to the time required for the forward propagation of the data through the network weights and activations. {For the XGBoost model, the estimation time is the time required for an input sample to pass through all decision trees in the ensemble and sum their outputs to generate the final prediction.} For the physics-based model, it is equal to the time for numerical integration of \eqref{thermal dynamics}-\eqref{electrical dynamics} for one time instant. All models have been tested on a Dell Precision 3660 desktop computer with Windows 11 Enterprise (Build 26200), equipped with an Intel Core i7-13700 processor (16 cores, 24 threads), 32 GB RAM, and an NVIDIA RTX A2000 GPU with 12 GB VRAM. GPU acceleration was enabled using NVIDIA driver version 573.44 and CUDA 12.8. The code was run in Python 3.13.5 using the PyTorch framework. Estimation time results show that KAN is the most efficient model with an estimation time of $0.0395 \mu s$, while MLP, RNN,  LSTM, and XGBoost models are roughly $1.4$,  {$21.7$,  $12.5$, and $20.4$} times slower than KAN, respectively.
Additionally, the physics-based model shows the highest latency at $1.3 \mu s$, approximately $32.9$ times slower than KAN. These results show that KAN offers the lowest computational latency, which makes it suitable for real-time and safety-critical applications. \\

\textbf{Criterion 3: Training RMSE.} The low training RMSE shows that all models fit the data well. LSTM reaches the lowest training RMSE of $0.0251$K with RNN following a close second with RMSE $0.0281$K. In contrast, KAN's training RMSE is $0.0345$K, which is sufficiently low while maintaining a small model size. {XGBoost achieves a slightly larger training RMSE of $0.0358$K than the KAN.}   Finally, MLP has the worst RMSEs of $0.0366$K among all models. This is expected as MLP is the least sophisticated model among the chosen baselines. \\

\textbf{Criterion 4: Validation RMSE.} The validation results are mostly commensurate with the training results. Even though the performance ranking of RNN and LSTM is flipped, their validation losses are still comparable at $0.0288$K and $0.0317$K, respectively. The KAN and XGBoost again exhibit similar validation RMSEs of \mbox{$0.0358$K} and $0.0354$K, respectively, which are very close to their corresponding training error.
Similarly,  MLP has the worst validation RMSE of $0.0581$K. As expected, all the validation errors are slightly higher than the training error. \\ 

\textbf{Criterion 5: Testing RMSE.} Similar to the validation results, the RNN achieves the lowest test RMSE of $0.0334$K, followed by LSTM with $0.0359$K. The proposed KAN-Therm model secures the next best testing RMSE of \mbox{$0.0477$K}.  The testing accuracy of the lightweight KAN-Therm model is acceptable since the difference in estimation error between the best-performing RNN and KAN-Therm model is $0.0143$K, which is comparable to the order of the sensor noise of $0.015$K. In contrast, XGBoost's testing performance is relatively worse with a larger RMSE of $0.0534$K. MLP achieves the least test accuracy and the largest RMSE of $0.0852$K.

\begin{figure*}[ht]
    \centering
    \includegraphics[height=0.4\textheight, keepaspectratio]{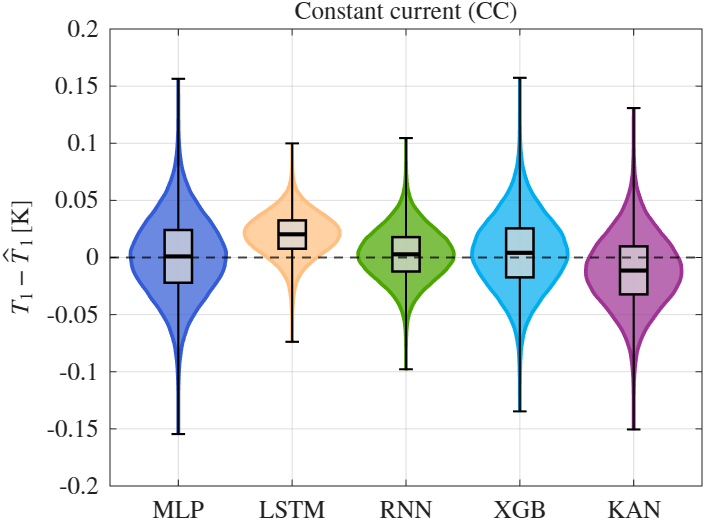}\\ %
    \includegraphics[height=0.4\textheight, keepaspectratio]{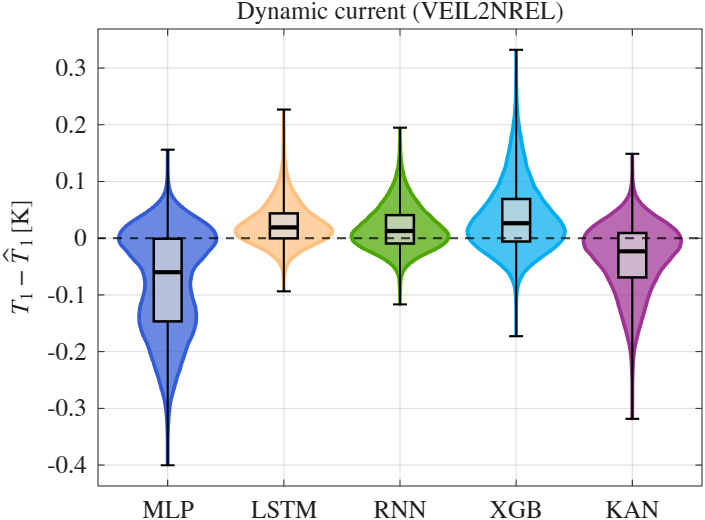}
    \caption{Plots showing the error distributions of core temperature estimation error under constant (top) and dynamic current (bottom) profiles.}
    \label{fig:violin2}
\end{figure*}

\textbf{Analysis of testing error distribution :}
\textbf{Analysis of testing error distribution:}
The testing RMSE presents a cumulative metric to evaluate the performance of each model. However, it is important to analyze the distribution of the temperature errors $T_1-\widehat{T}_1$ of the testing dataset for each of the five models. 
Fig.~\ref{fig:violin2} shows this error distribution using violin plots, whose width indicates how frequently the error value occurs.  This figure also presents the box plots for the interquartile range, the median error line, and the maximum-minimum lines for the error range. The top plot here corresponds to the constant current testing and the bottom plot is for the VEIL2NREL testing. In the VEIL2NREL testing scenario, we test the performance of each model for an unseen dynamic profile. This task is significantly more challenging than estimating for a constant unseen profile. Thus, we observe that the error ranges are significantly higher for the VEIL2NREL testing scenario compared to the constant current testing for all the methods.

Thus, our extensive analysis of the models concludes that the proposed KAN-Therm model achieves comparable and practical test accuracy with greater efficiency and lower parameter count, which makes it suitable for scalability and an excellent choice for real-time applications on resource-limited physical BMS.

Table~\ref{tab:per_compSumm} summarizes the comparison results and  highlights the trade-off between the estimation accuracy and the model complexity. Although RNN achieves the best testing accuracy, it requires nearly five times more parameters than KAN. In contrast, KAN exhibits  larger RMSE compared to RNN while remaining significantly smaller and faster. Despite this larger RMSE, the relative error of KAN is approximately $0.0125\%$, which indicates that core temperature estimation from KAN can be reliably utilized for downstream applications. Therefore, KAN provides a strong balance between accuracy and computational efficiency for battery management systems with limited memory and computational resources.

\section{Conclusion} \label{conclusion}
In this work, we propose a lightweight, computationally efficient data-driven battery thermal model based on the KAN architecture to accurately estimate the battery core temperature. This proposed KAN-therm model leverages KAN's learnable activation functions to efficiently learn the nonlinear relationship between the battery core temperature and its surface temperature, coolant temperature, coolant flow rate, and current. Training, validation, and testing of the model are conducted using an extensive dataset generated for both constant and dynamic current profiles. The hyperparameters of the final KAN-Therm model are identified through an extensive grid search to ensure the best performance in terms of accuracy. Although the best-performing RNN model achieves slightly lower core temperature estimation error, its difference from KAN-Therm in testing RMSE is only $0.0143$K. Despite the minor loss in accuracy, KAN significantly reduces the parameter count by $79.3$\% and improves estimation time by $21.7$ times in comparison to RNN. Similarly, although LSTM achieves slightly lower estimation error than KAN-Therm, it requires a significantly larger network and $12.5$ times longer estimation time. Compared with XGBoost, KAN-Therm achieves lower testing RMSE while reducing the parameter count by $99.6$\% and improving estimation time by $20.4$ times. While the MLP model performed relatively faster than RNN and LSTM, it underperformed the KAN-Therm model in accuracy, estimation time, and network complexity. Overall, these results show that the KAN-therm model achieves an effective balance between accuracy and computational resources, highlighting its potential applicability in a low-resource battery management system hardware.





\bibstyle{arxiv}
\bibliography{ref1.bib}
\end{document}